\documentclass[11pt]{article}
\pdfoutput = 1

\usepackage[utf8]{inputenc}
\usepackage{color,graphicx}
\usepackage{amsmath}
\usepackage{verbatim}
\usepackage{amssymb}
\usepackage{physics}
\usepackage{amsfonts}
\usepackage{cite}
\usepackage{array}
\usepackage{setspace}
\usepackage{float}
\usepackage{url}
\usepackage{tikz}
\usepackage{graphicx}
\usepackage{fancybox}
\usepackage{tensor}
\usepackage{tikz}
\usepackage{cite}
\usepackage{caption}
\usepackage{subcaption}
\allowdisplaybreaks

\usepackage[margin = 2.2cm]{geometry}
\usepackage[ragged]{footmisc}
\usepackage{hyperref}
\hypersetup{
    colorlinks,
    citecolor=blue,
    filecolor=black,
    linkcolor=blue,
    urlcolor=blue,
    linktocpage=true
}

\def\Tr{\operatorname{Tr}}

\newcommand{\be}{\begin{equation}}
\newcommand{\ee}{\end{equation}}
\newcommand{\bea}{\begin{align}}
\newcommand{\eea}{\end{align}}
\newcommand{\bi}{\begin{itemize}}
\newcommand{\ei}{\end{itemize}}

\newcommand{\jt}{{\text{JT}}}
\newcommand{\lr}[1]{\left( #1 \right)}

\def\t{\text}

\def\rb{\rangle}
\def\lb{\langle}
\def\hh{\Psi_{\t{HH}}}
\def\hhb{\Psi_{\Sigma_B}}

\numberwithin{equation}{section}

\begin{document}

\thispagestyle{empty}
\begin{center}
\vspace*{.4cm}
     {\LARGE \bf 
Closed universes, factorization, and ensemble averaging}
    
    \vspace{0.4in}
    {\bf Mykhaylo Usatyuk, Ying Zhao}

    \vspace{0.4in}
{Kavli Institute for Theoretical Physics, Santa Barbara, CA 93106, USA}
    \vspace{0.1in}
    
    {\tt musatyuk@kitp.ucsb.edu,\ \ \ zhaoying@kitp.ucsb.edu}
\end{center}

\vspace{0.4in}
\begin{abstract}
We study closed universes in holographic theories of quantum gravity. We argue that within any fixed theory, factorization implies there is one unique closed universe state. The wave function of any state that can be prepared by the path integral is proportional to the Hartle-Hawking wave function. This unique wave function depends on the properties of the underlying holographic theory such as the energy spectrum. We show these properties explicitly in JT gravity, which is known to be dual to an ensemble of random Hamiltonians. For each member of the ensemble, the corresponding wave function is erratic as a result of the spectrum being chaotic. After ensemble averaging, we obtain smooth semi-classical wave functions as well as different closed universe states.

\end{abstract}

\pagebreak
\setcounter{page}{1}
\tableofcontents

\newpage

\section{Introduction}

Gauge/gravity duality is a powerful framework to study quantum gravity \cite{Maldacena:1997re,Gubser:1998bc,Witten:1998qj}. A lot of progress has been made in understanding quantum gravity in asymptotically AdS spacetimes through the AdS/CFT correspondence. However, the universe we live in has a positive cosmological constant and may not have an asymptotic boundary. The understanding of closed universes without boundaries has been limited. 



In \cite{Usatyuk:2024mzs} we studied simple models of closed universes in two dimensional gravity and pointed out various puzzling features. In particular, semi-classically there appear to be many quantum gravity states encoding rich physics, but once non-perturbative effects are included there is only a single closed universe state \cite{Coleman:1988cy,Maldacena:2004rf,Penington:2019kki,Marolf:2020xie,Marolf:2020rpm,Susskind:2021omt,McNamara:2020uza}. In this work we continue to study these issues. The approach we will take is to begin with a well-defined holographic theory dual to a certain bulk theory, and we will study closed universes in this theory. In section \ref{sec:general} we point out the following general features.
\begin{itemize}
    \item With a single holographic theory:
    \begin{itemize}
        \item Factorization implies there is only one closed universe state. Any state that can be prepared by the path integral has a wave function proportional to the Hartle-Hawking wave function.
        \item The data of the boundary theory, such as the spectrum, is intricately encoded in the unique wave function. For chaotic holographic theories, we expect the wave function will be erratic.
        \item A more refined argument using factorization shows that the unique wave function, in a certain sense, contains trivial information. All states we can input into the wave function are equivalent.
    \end{itemize}

    \item With an ensemble average over theories:
     \begin{itemize}
        \item Factorization is no longer true. Distinct closed universe states can be prepared by specifying different boundary conditions.
        \item The wave functions contain non-trivial information. There are a large number of orthogonal states that can be used as inputs for the wave functions.
        \item The wave functions are smooth functions of the inputs, and are well approximated by a sum over semi-classical geometries.
    \end{itemize}
\end{itemize}
We will be able to show some of the features listed above hold generally, regardless of the specific model being studied. For other features, we can only rigorously show them in simple models. Nevertheless, we expect these are general lessons that hold more broadly for holographic theories.


In section \ref{sec:JT} we consider the concrete model of 2d JT gravity. Pure JT gravity is known to be dual to an ensemble average of Hamiltonians \cite{Saad:2019lba}. We will restrict to a single Hamiltonian by considering a specific deformation of the theory\cite{Blommaert:2021fob}. We will demonstrate all of the points listed above in this model, and find an explicit expression for the unique wavefunction in terms of the boundary data. One surprising feature we will see is that even though we have argued the closed universe wave function within a fixed single theory is trivial, it nevertheless shares some qualitative similarities with the wave function of the averaged theory.

Closed universes having a single quantum state in quantum gravity is an old claim \cite{Coleman:1988cy,Giddings:1988cx,Giddings:1988wv,Maldacena:2004rf} that has received renewed interest in recent years \cite{Penington:2019kki,Marolf:2020xie,McNamara:2020uza}, and in section \ref{sec:general} we review these claims. In \cite{Coleman:1988cy,Giddings:1988cx,Giddings:1988wv} it was pointed out that wormholes can play a fundamental role in closed cosmologies. \cite{Maldacena:2004rf} gave explicit constructions of such wormholes in the framework of AdS/CFT. The implications of these claims was studied in detail in \cite{McInnes:2004ci,McInnes:2004nx}. \cite{Penington:2019kki,Marolf:2020xie,McNamara:2020uza}
gave a more modern perspectives which is closer to our approach. The new material in this work is largely contained in section \ref{sec:JT} where we studied various features in a concrete toy model and explicitly obtained both erratic and smooth wave functions with and without ensemble averaging.

In section \ref{sec:discussion} we point out unanswered questions and future directions.

\section{Factorization and the uniqueness of closed universe state}
\label{sec:general}

\subsection{Preparing states of closed universes}

In this section we will assume we are starting with a single holographic boundary theory dual to a bulk gravitational theory, and that the bulk theory is specified by a bulk gravitational path integral. We are interested in understanding the bulk theory on spatial slices that do not have any asymptotic boundaries. In such cases, the connection between the bulk and boundary theory is not well understood. 

We begin by constructing quantum states of closed universes on a compact Cauchy slice $\Sigma$. We allow for $\Sigma$ to be the union of multiple disconnected compact slices. States are specified by a wave function of the induced metric $h$ on the slice, along with other potential fields. The most obvious state we can construct is the Hartle-Hawking state $\hh[\Sigma]$, where we perform a path integral over all manifolds bounded by a slice $\Sigma$ with data $h$ on the slice \cite{HartleHawking83}
\be
\hh[\Sigma] = ~\raisebox{-0.6cm}{\includegraphics[width=2cm]{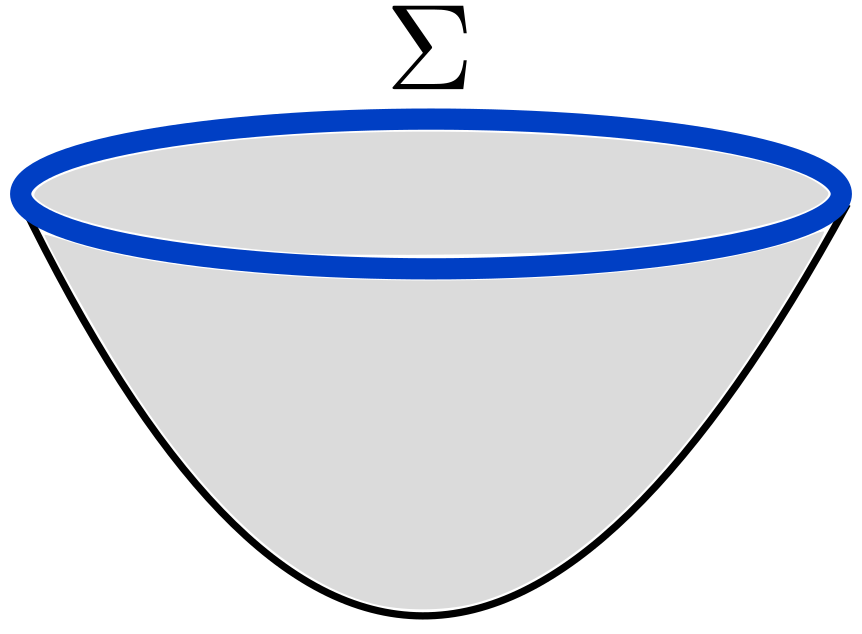}} ~~ + \ldots
\ee
The Hartle-Hawking prescription has been studied in AdS space in a variety of works \cite{Maldacena:2004rf,McInnes:2004ci,McInnes:2004nx,Chen:2020tes,Navarro-Salas:1992bwd,Marolf:2020xie,Marolf:2020rpm,Iliesiu:2020zld,Maldacena:2019cbz,Chen:2020tes,VanRaamsdonk:2020tlr,Cooper:2018cmb,VanRaamsdonk:2021qgv,Antonini:2022blk,Antonini:2022ptt,Sahu:2023fbx,Antonini:2023hdh}. However, this is not the only choice. We can construct other states by specifying additional boundaries with boundary conditions denoted by $\Sigma_B$. Then $\Psi_{\Sigma_B}[\Sigma]$ is given by the gravitational path integral over all manifolds bounded by $\Sigma_B$ and the spatial slice $\Sigma$ with induced metric $h$. In this notation the Hartle-Hawking state is specified by the empty set $\Sigma_B = \varnothing$. Naively, by specifying different boundary conditions we obtain different wave functions $\Psi_{\Sigma_B}$, and hence different closed universe states. Let's examine this more closely. Consider the case where $\Sigma_B$ is an asymptotic boundary circle with length $\beta$, which computes $Z(\beta) = \Tr(e^{-\beta H})$ in the corresponding boundary theory. From the bulk path integral perspective it appears we have produced a new wave function since there exist bulk manifolds that connect $\Sigma_B$ to $\Sigma$, pictorially
\be
\Psi_\beta[\Sigma] = \raisebox{-1.7cm}{\includegraphics[width=6cm]{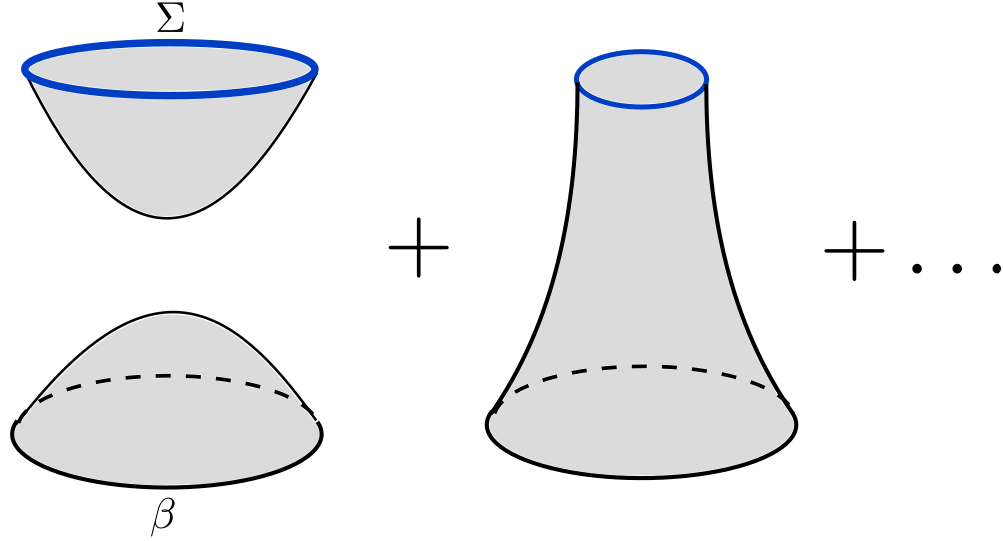}} \,.
\ee
However, since the boundary theory is a well-defined quantum mechanical theory, $Z(\beta)$ is simply a number. This implies that for arbitrary spatial slices $\Sigma$ we must have
\begin{align}
    \Psi_{\beta}[\Sigma] = \hh[\Sigma] \times Z(\beta)\,.
\end{align}
Another way to state the above is that the bulk path integral that computes the cosmological state with additional asymptotic boundary conditions must factorize, which is a variant of the standard factorization problem in holography\cite{Maldacena:2004rf} applied to the case of computing the closed universe wave function. One may worry that factorization may not hold here because $\Sigma$ is a finite boundary. However, from the perspective of the boundary quantum theory, the inclusion of asymptotic boundaries $\Sigma_B$ computes an independently well defined quantity, in this case $Z(\beta)$. As a result, we must have that asymptotic boundaries factorize from the finite boundary $\Sigma$.

This argument applies to arbitrary asymptotic boundary conditions $\Sigma_{B}$. The boundary theory on asymptotic boundaries $\Sigma_B$ simply computes an independently well defined boundary quantity $Z[\Sigma_B]$. By factorization within a single theory we have
\begin{align}
\label{factorization}
    \Psi_{\Sigma_B}[\Sigma] =\hh[\Sigma] \times Z[\Sigma_B]\,.
\end{align}
We conclude that for any fixed holographic theory there exists one closed universe state: the Hartle-Hawking state. All other states prepared through the bulk path integral have wave functions proportional to the Hartle-Hawking wave function. In \cite{Usatyuk:2024mzs} it was argued that in a fixed theory the states of closed universes are numbers. From the above argument these numbers are the prefactors in front of the Hartle-Hawking wave function, given by $Z[\Sigma_B]$ if the state is prepared with asymptotic boundary conditions $\Sigma_B$.

We again emphasize that $\Sigma$ can be a union of disconnected compact manifolds. $\hh[\Sigma]$ gives the unique closed universe wave function with support on any number of disconnected universes.

\subsection{Hartle-Hawking wave function in a single theory}

A natural question to ask is how to compute the unique Hartle-Hawking wave function given a well defined holographic theory. There are two approaches one can take. 

From the bulk perspective, if we know the exact bulk theory we can directly perform the gravitational path integral following the procedure given by Hartle and Hawking \cite{HartleHawking83}. Additionally, given the complete description of the bulk path integral we should find that factorization \eqref{factorization} automatically holds.

From the boundary perspective, we would need to understand how the closed universe is encoded in the boundary data. As the spatial slice $\Sigma$ does not have a boundary, one may worry that the boundary theory knows nothing about the closed universe. We find simple theories where this is not the case. We will show in the next section that boundary data, such as the energy spectrum of the Hamiltonian, is intricately encoded in the exact Hartle-Hawking wave function. Going in reverse, given boundary data we can reconstruct the Hartle-Hawking wave function for closed universes. 


A natural guess is that these properties extend more generally beyond the models we consider. If this is the case we can infer some general properties of the wave function in a single theory. Holographic theories with bulk duals that resemble Einstein gravity are chaotic \cite{Cotler:2016fpe}. Assuming the wave function depends on the details of the spectrum, we should expect to find that $\hh[\Sigma]$ is erratic for a single theory. However, performing an ensemble average can smooth it out and give us wave functions that look smooth and semi-classical.


\subsubsection*{Trivial wave function}

We have concluded that given a single holographic theory, factorization implies there is a unique wave function of any states prepared by path integral. Naively this wave function carries interesting physical information, for example it can tell us the relative amplitude to find a closed universe with a particular field configuration on a Cauchy slice. This turns out to be incorrect.

We can take two states specified by different field configurations $\lr{\Sigma_1 , h_1}$, $\lr{\Sigma_2 , h_2}$ and consider their inner product. The inner product is defined by performing a bulk path integral over all manifolds bounded by $\Sigma_1$ and $\Sigma_2$ with their specified data. Since the bulk path integral factorizes for a single theory, we immediately have
\begin{align}\label{eqn:inner_product}
    \bra{\Sigma_1}\ket{\Sigma_2} = \hh[\Sigma_1 \cup \Sigma_2] = \hh[\Sigma_1]\hh[\Sigma_2]\,.
\end{align}
Here we should emphasize that we are assuming the stronger condition that finite boundaries also factorize in a single theory, see section \ref{sec:discussion} for discussions. To check whether states are linearly independent and correspond to distinct states in the Hilbert space, we can construct their Gram matrix
\begin{align}
\label{eq:rank_inner_product} G_{ij}\equiv\bra{\Sigma_i}\ket{\Sigma_j} = \hh[\Sigma_i]\hh[\Sigma_j]\,.
\end{align}
The rank of the Gram matrix tells us the number of linearly independent states. From equation \eqref{eq:rank_inner_product} one immediately sees that the gram matrix has rank one. This stems from the fact that with a factorizing inner product as above, the inner product between any two states cannot be zero unless one of the states is null. This implies that slices with different metrics/field content correspond to the same physical state, and in this sense the unique wave function in a single theory appears to be trivial.

\subsection{Ensemble averaging}

Let's briefly summarize how ensemble averaging over boundary theories can change the preceding story. It has been speculated that ensemble averaging may play an important role in cosmology, see \cite{Banks:2002wr,Maldacena:2019cbz,Chen:2020tes,VanRaamsdonk:2020tlr,Susskind:2021omt}. Our argument for a single closed universe state crucially relies on the property of factorization. With an ensemble average of theories, factorization no longer holds, and we have the following consequences:
\begin{itemize}
    \item After ensemble averaging the gram matrix defined in \eqref{eq:rank_inner_product} no longer has rank one, which implies different field configurations can correspond to physically distinct closed universe states. 
    \item Distinct states with different wave functions can be prepared by specifying different asymptotic boundary conditions.
    \item The closed universe wave functions can be smooth and behave semi-classically. 
\end{itemize}

\section{From matrix integrals to closed universes}
\label{sec:JT}
In this section we will work with JT gravity which is dual to a matrix integral. We will explicitly illustrate the statements made in the previous section. 

\subsection{Review of JT gravity, matrix integral, and a factorizing theory}
The JT gravity action in Euclidean signature is given by \cite{Teitelboim:1983ux,Jackiw:1984je,MSY16}\footnote{Throughout this section we set $8\pi G_N = 1$.}
\be
I_\jt[g,\Phi]=-S_0 \chi -\frac{1}{2}\int \sqrt{g}\Phi(R+2)-\int_{\text{bdy}} \sqrt{h} \Phi_b(K-1) \,,
\ee
in terms of the dilaton $\Phi$ and metric $g$. The term $ S_0 \chi$ suppresses higher topologies by the Euler characteristic. We have included a boundary term so that good asymptotic boundary conditions consist of fixing the dilaton $\Phi|_{\partial M}$ and the metric along the boundary.

Pure JT gravity is dual to an integral over random matrices \cite{Saad:2019lba}. More precisely, the duality is
\be \label{eqn:JTduality}
\int d \mu(H) \Tr\lr{e^{-\beta_1 H}} \ldots \Tr\lr{e^{-\beta_n H}} = \int_{\t{b.c.}} \mathcal{D} g \mathcal{D} \Phi e^{-I_\jt [g, \Phi]} \,.
\ee
On the left-hand side we have an integral over Hermitian matrices $H$ with a particular measure $\mu(H)$.\footnote{The double scaling limit must be taken where the size of the matrices is sent to infinity while we zoom into the end of the spectrum. The measure is suitably tuned so that amplitudes computed in the matrix integral match order by order in the topological expansion to JT gravity amplitudes.}  The matrix $H$ is to be interpreted as the Hamiltonian of a boundary quantum mechanical theory. 
Operator insertions in the matrix integral, such as $\Tr\lr{e^{-\beta H}}$, map to gravitational boundary conditions such as an asymptotic boundary of length $\beta$ \cite{Saad:2019lba,Goel:2020yxl}.


\subsubsection*{A factorizing model} 

While pure JT gravity is dual to a matrix integral, we are interested in studying a bulk theory dual to a single quantum mechanical theory with fixed Hamiltonian. Following \cite{Blommaert:2021fob,Blommaert:2022ucs}, we can consider a deformed bulk theory with the following action:
\be
\label{eq:action_deformed}
I[g,\Phi] = I_\jt[g, \Phi] - \underbrace{\int_0^\infty d b b \mathcal{O}(b)  Z_{H_0}(b)}_{\t{gives discrete spectrum}} + \underbrace{\frac{1}{2}\int_0^\infty b db \mathcal{O}(b) \mathcal{O}(b)}_{\substack{ \t{gives bulk factorization}}}\,.
\ee
The operator $\mathcal{O}(b)$ when inserted in the path integral creates a geodesic boundary of length $b$ on which surfaces can end.\footnote{This operator has an expression in terms of dilaton gravity variables given by $\mathcal{O}(b) \Leftrightarrow e^{-S_0} \int_{\mathcal{M}} d^2 x \sqrt{g(x)} e^{-2\pi \Phi(x)} \cos \lr{b \Phi(x)}$. However, it is more convenient to express it simply as $\mathcal{O}(b)$ when performing the series expansion of the deformation.} The function $Z_{H_0}(b)$ is given by
\be
\label{eq:def_Zb}
Z_{H_0}(b)=\sum_{i=1}^N \frac{2}{b} \cos \left(b \sqrt{E_i}\right)-\int_0^{\infty} d E \rho_0(E) \frac{2}{b} \cos \left(b \sqrt{E} \right)\,,
\ee
where the energies $E_i$ are the eigenvalues of a Hamiltonian $H_0$ and $\rho_0(E)=\sinh \lr{2\pi \sqrt{E}}$ is the disk density of states of pure JT. The second deformation term is non-local in spacetime, and introduces a condensate of correlated geodesic boundaries. The role of this deformation is to factorize the bulk theory by canceling all connected wormhole contributions order by order in the genus expansion. It was argued that this deformed bulk theory is dual to a single fixed boundary Hamiltonian $H_0$ whose discrete spectrum is encoded in the function $Z_{H_0}(b)$.

Amplitudes in the theory are defined by first defining a set of gravitational boundary conditions, expanding the deformations in a perturbative series, and evaluating the amplitude at each order in the expansion in the pure JT theory. The gravitational boundary conditions can now be connected to the geodesic boundaries generated by $\mathcal{O}(b)$ at each order in the expansion. It's simplest to represent the deformation terms through pictorial rules for the bulk path integral\footnote{The interaction terms put into the action are renormalized interaction vertices where all self-interactions have already been defined to be re-summed. Thus the interaction vertices that are brought down in the series expansion cannot connect to each other to create degenerate surfaces.}
\begin{align}
&\t{Discrete spectrum interaction:} \qquad \qquad \int_0^\infty d b b \mathcal{O}(b)  Z_{H_0}(b)  \Leftrightarrow ~~ \raisebox{-.75cm}{\includegraphics[width=1.1cm]{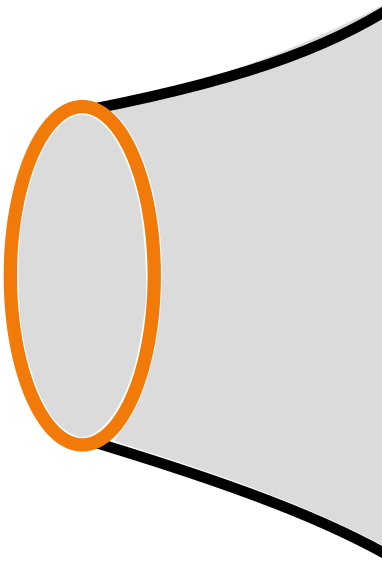}}\\
&\t{Factorization branes:} \qquad \qquad -\frac{1}{2}\int_0^\infty d b b \mathcal{O}(b) \mathcal{O}(b) \Leftrightarrow ~~ \raisebox{-.75cm}{\includegraphics[width=2.2cm]{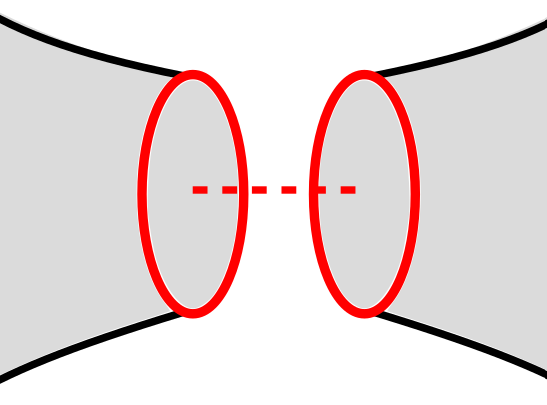}}
\end{align}
The factorizing branes are pictured linked since their geodesic lengths $b$ are the same.\footnote{For the rest of the paper, we will use orange and red circles to represent the geodesic boundaries $\mathcal{O}(b)$ generated by the respective interaction terms.} For the convenience of the readers, we briefly explain how factorization works in this model in appendix \ref{app:factorization}. Readers who are interested in more details can look at the original paper \cite{Blommaert:2021fob}.  

\subsection{Closed universes in JT gravity}

We first look at closed universes in pure JT gravity \cite{Louis-Martinez:1993bge,Navarro-Salas:1992bwd,Henneaux:1985nw,Usatyuk:2024mzs}. For related recent works, see \cite{Marolf:2020xie,Marolf:2020rpm,Iliesiu:2020zld,Maldacena:2019cbz,Chen:2020tes,VanRaamsdonk:2020tlr,Cooper:2018cmb,VanRaamsdonk:2021qgv,Antonini:2022blk,Antonini:2022ptt,Sahu:2023fbx,Antonini:2023hdh}. The unique classical solution in Lorentzian signature with compact spatial slices $S^1$ gives a big bang/big crunch cosmology\footnote{This solution has conical singularities at $t=\pm \frac{\pi}{2}$ instead of genuine curvature singularities as in higher dimensional big bang/big crunch cosmologies.}
\be
ds^2 = - dt^2 + b^2 \cos^2(t) d \sigma^2, \qquad \Phi(t) = \phi_c \sin (t)\,.
\ee
The spatial coordinate is periodic $\sigma \sim \sigma + 1$ and time runs from $t \in (-\frac{\pi}{2},\frac{\pi}{2})$. The universe exists for a finite amount of time, and reaches a maximal spatial size $b$ at time $t=0$ with the slice being a geodesic. 


To discuss the quantum theory of these geometries we can compute the Hartle-Hawking wave function \cite{HartleHawking83}. We integrate over all Euclidean geometries ending on a slice with extrinsic curvature $K=0$ and geodesic size $b$ (blue circle in \eqref{eqn:HH_PureJT}) . In JT gravity this wave function has a simple expression\cite{Saad:2019lba}\footnote{Integrating the dilaton over a complex contour restricts the integral to $R=-2$ geometries, so we must sum over constant negative curvature geometries.}
\be \label{eqn:HH_PureJT}
\hh(b) = ~~~ \raisebox{-1.5cm}{\includegraphics[width=2cm]{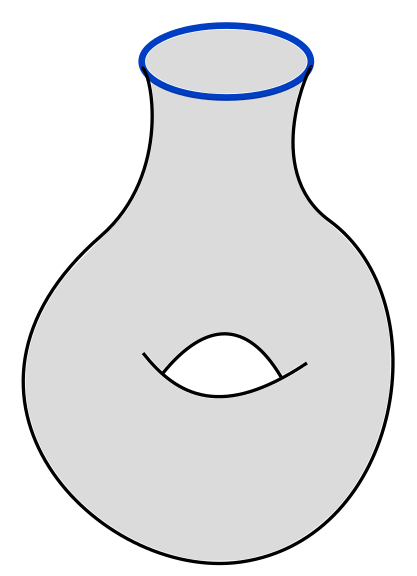}} + \ldots ~~ = \sum_{g=1}^\infty e^{S_0 \chi} V_g(b)\,.
\ee
given by a sum over hyperbolic manifolds bounded by a geodesic boundary $b$, with each genus $g$ geometry contributing a Weil-Petersson volume weighed by the Euler-characteristic.\footnote{The knowledge of Weil-Petersson volumes is not necessary for this work, but closed form expressions are known.} The final result gives an amplitude to find a universe of size $b$.

The Hartle-Hawking prescription is one choice of asymptotic boundary condition, namely no boundaries, from which we obtain a state. We can obtain other states by demanding the insertion of additional boundaries $\Sigma_B$ in the Euclidean path integral. This would prepare a different closed universe state depending on $\Sigma_B$. For example, we can insert an asymptotic boundary of length $\beta$, and integrate over all bulk geometries ending on the geodesic slice $b$ and asymptotic boundary\footnote{Explicit expressions for all these amplitudes can be readily found in \cite{Saad:2019lba}.}
\begin{align} 
\hhb (b)  = ~~~ \raisebox{-1.5cm}{\includegraphics[width=4.5cm]{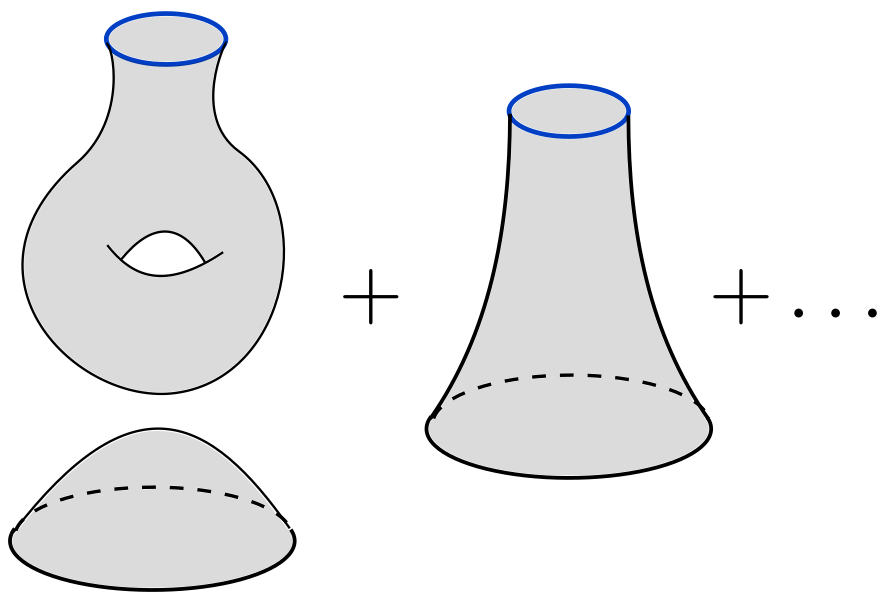}} ~~ =  V_{1}(b) Z_{\t{disk}}(\beta) + Z_{\t{trumpet}} (\beta, b)   +\ldots\,.
\end{align}
The bulk manifold connecting the $\Sigma_B$ boundary and the $b$ boundary is a version of the factorization problem. In this case it is not a paradox since we are explicitly averaging over boundary theories. The lack of factorization allows us to construct many closed universe states $\hhb (b)$ by varying the boundary conditions $\Sigma_B$. We shall now explicitly see this is no longer the case in a theory that displays bulk factorization.

\subsection{The unique closed universe wave function in a fixed theory}
Let's us now consider the wave function of closed universes in the single fixed theory. To obtain the wave function we insert a geodesic boundary of size $b$ and perform the path integral in a perturbative series in genus and in the interactions. The Hartle-Hawking wave function is given by
\begin{align}
\label{eq:HH_1}
\hh (b) = ~~\raisebox{-1.2cm}{\includegraphics[width=6cm]{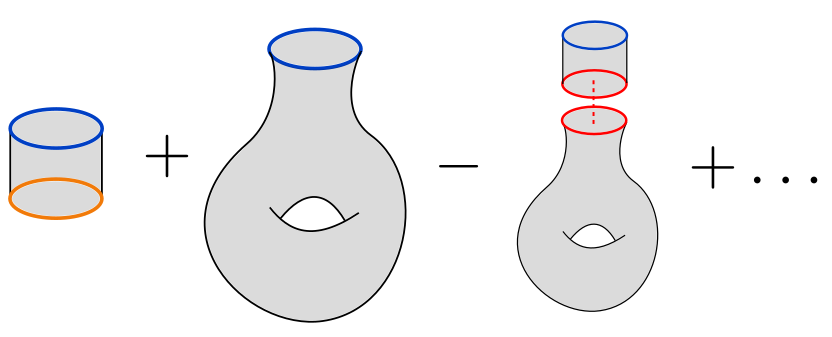}}
\end{align}
It is not difficult to see see that all higher genus geometries will be cancelled by the factorizing branes. Furthermore, the only geometric contribution that survives is the one where the $Z_{H_0}(b)\mathcal{O}(b)$ vertex connects to our geodesic boundary. Such a geometry is a degenerate cylinder of zero size sandwiched between geodesic boundaries of lengths $b$ and $b'$. The amplitude for this degenerate cylinder is given by $\lb b | b' \rb = \frac{1}{b} \delta(b-b')$. The Hartle-Hawking wave function is thus given by
\begin{align}
\hh (b) &= \int_0^\infty db' b' Z_{H_0}(b') \times \frac{1}{b} \delta(b-b')=Z_{H_0}(b)\,.\label{eq:HH_wavefunction} 
\end{align}
We can attempt to modify this closed universe wave function by inserting additional asymptotic boundaries. Let's see how the condensate of factorization branes ensures this does not work. The simplest geometric contribution is given by connecting the asymptotic boundary to the geodesic boundary,
\be
\label{cancellation}
\hhb(b) \supset ~~\raisebox{-1.2cm}{\includegraphics[width=3cm]{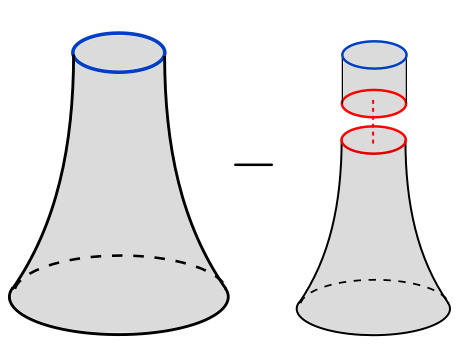}} =0 \,.
\ee
However, as in \eqref{cancellation}, the factorizing branes give an identical geometry with a relative sign, canceling this contribution. All geometries connecting the asymptotic boundary $\Sigma_B$ and the geodesic circle $b$ cancel for similar reasons. The only geometries that end up surviving are 
the following:
\be
\hhb(b) = ~\raisebox{-.7cm}{\includegraphics[width=6cm]{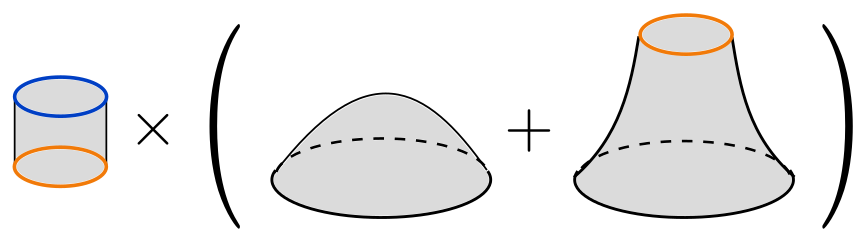}}  ~ = \hh (b) \times Z(\beta) \,.
\ee
This is precisely the answer we should expect from a factorizing theory as discussed in section \ref{sec:general}. Thus we see we cannot change the state of the closed universe by changing the boundary conditions. This same conclusion continues to hold if we insert any other type of boundary conditions into the theory, since the factorizing branes would kill off any kind of wormhole contributions. 

So far we have performed a bulk computation and found that in the factorizing theory all closed universe states have wave functions proportional to the Hartle-Hawking wave function
\be
\hhb(b) = \hh(b) \times (\t{constant})\,.
\ee

Combining equations \eqref{eq:def_Zb} and \eqref{eq:HH_wavefunction}, we have
\be
\label{eq:HH_dictionary}
\hh (b) = \sum_{i=1}^N \frac{2}{b} \cos \left(b E_i^{1 / 2}\right)-\int_0^{\infty} d E \rho_0(E) \frac{2}{b} \cos \left(b E^{1 / 2}\right)\,.
\ee
This equation \eqref{eq:HH_dictionary} directly relates the unique wave function of the closed universe to the properties of the dual boundary theory. We point out a few aspects of this relation.
\bi
\item {The Hartle-Hawking wave function is fixed by the data of the boundary dual.} In this case, the boundary data consist of the energy eigenvalues $E_i$ of the boundary Hamiltonian.
\item {The wave function is erratic for a single member of the ensemble.} For a particular draw of the random matrix $H_0$ we can plot the wave function:\footnote{The numerics were obtained by keeping the square root edge of the spectrum of a random matrix, and cutting off the integral on the disk subtraction at the largest eigenvalue of the random matrix. The matrix ensemble was chosen to be the gaussian unitary ensemble instead of the JT ensemble for simplicity.}
\begin{figure}[H]
    \centering
    \includegraphics[width=7cm]{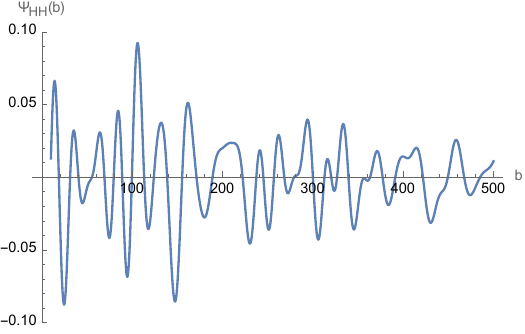}
    \caption{The Hartle-Hawking wave function $\hh(b)$ for a single draw of a random matrix $H_0$. Since the spectrum is chaotic the resulting wave function is highly oscillatory and sensitive to the eigenvalues. In this plot we take the first 100 eigenvalues out of a $10^4$ sized Hermitian matrix, with typical level spacing $\sim 10^{-3}$ near the edge.}
    \label{fig:enter-label}
\end{figure}
The erratic behavior comes from the fact that the spectrum for a random Hamiltonian $H_0$ is chaotic.
\item {We can extract boundary data if given the exact Hartle-Hawking wave function.}
\ei

A priori it is unclear what (if any) information closed universes carry about the dual theory on the boundary. After all, closed universes themselves do not have an asymptotic boundary. We see in this theory that the Hartle-Hawking wave function encodes boundary data in a particularly straightforward way. Given the above wave function \eqref{eq:HH_dictionary}, we can extract the eigenvalues of the boundary Hamiltonian by performing integral transforms. For example, we have
\be
\frac{1}{\pi}\int_0^\infty db \hspace{.03cm}b \hh(b)\cos(b \sqrt{E}) = \sum_{i=1}^N \qty[\delta(\sqrt{E}-\sqrt{E_i})+\delta(\sqrt{E}+\sqrt{E_i})]-\rho_0(E)\,.
\ee
In section \ref{sec:general} we also argued that this unique wave function is trivial in the sense that all inputs of the wave function are the same, i.e., different $b$'s correspond to the same state. In the current model this is obvious, as from \eqref{cancellation} overlaps factorize and we have $\lb b |b' \rb=\hh(b) \hh(b')$ with a fixed Hamiltonian $H_0$ and \eqref{eq:rank_inner_product} manifestly holds. Said another way, $b,b'$ are not distinct states.

\subsection{Ensemble averaging: Going back to JT gravity with smooth wave functions}
We now briefly comment on the boundary matrix integral interpretation of these wave functions. The matrix operator that corresponds to inserting a geodesic boundary in gravity is given by\cite{Goel:2020yxl,Blommaert:2021fob}\footnote{The subtraction of the second term gets rid of the disk level density of states contribution when the operator is integrated over all random matrices. This is needed because in gravity a hyperbolic disk with geodesic boundary does not exist, and the first geometry contributing to the amplitude exists at genus one. }
\be \label{eqn:matrixoperator_O}
\mathcal{O}(b) \Longleftrightarrow  \frac{2}{b} \Tr \cos \left(b \sqrt{H}\right)-\int_0^{\infty} d E \rho_0(E) \frac{2}{b} \cos \left(b E^{1 / 2}\right)\,.
\ee
The expectation value of this operator in the boundary theory is thus the Hartle-Hawking wave function in the bulk. The deformations added to the bulk action in \eqref{eq:action_deformed} map to a deformation of the matrix integral potential \eqref{eqn:JTduality}, and localize the integral onto a single Hamiltonian $H_0$ with eigenvalues $\lr{E_1, \ldots, E_N}$ set by the parameter $Z_{H_0}$ \cite{Blommaert:2021fob}. This reproduces the Hartle-Hawking wave function as computed from the bulk earlier. 

The unique wave function with a fixed Hamiltonian has some undesirable features: it is erratic, does not appear to carry non-trivial information, and it is unclear how to do quantum mechanics with one state. These issues can be solved by ensemble averaging over Hamiltonians to get back to pure JT. The expectation of $\mathcal{O}(b)$ for each Hamiltonian will be erratic, but if we average over the Hamiltonians we recover a smooth answer
\be \label{eqn:AveragedHH}
\int d \mu(H) \mathcal{O}(b) \Longleftrightarrow  \hh(b) = \sum_{g=1}^\infty e^{S_0 \chi} V_g(b)\,.
\ee
Furthermore, after averaging we can now prepare different cosmological states, since the bulk path integral no longer factorizes.\footnote{If wormhole corrections are included one would of course see large corrections/fluctuations in observables. Averaging only helps in constructing non-trivial closed universe Hilbert spaces if we modify the inner product to not include wormhole corrections. This seems unnatural from the perspective of recent developments in black hole physics, but maybe the rules need to be modified for cosmology.} For example, we have that the overlap between different $b$ basis states is now $\bra{b}\ket{b'} = \frac{1}{b}\delta(b-b')$, whereas before it was the product of two random numbers given by \eqref{eq:HH_dictionary}.

One natural question is whether the wave function of a single theory \eqref{eq:HH_dictionary} knows anything about the averaged wave function \eqref{eqn:AveragedHH}. The answer is yes but in a subtle way. The full answer in \eqref{eqn:AveragedHH} will grow as a function of $b$. One may wonder if we can see this growth from the wave function of a single theory. At least qualitatively, the answer is yes:
\begin{figure}[H]
\centering
     \begin{subfigure}[b]{0.4\textwidth}
    \centering
        \includegraphics[width=\textwidth]{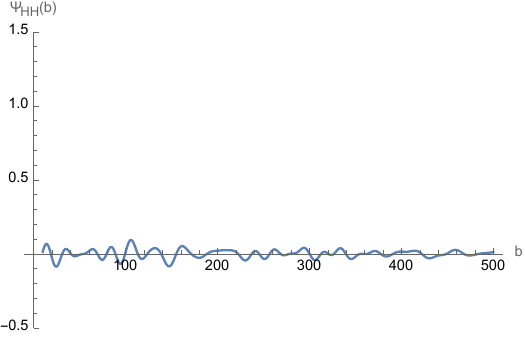}
\caption{Typical draw of $H_0$}
\label{a}
\end{subfigure}
\quad\quad
\begin{subfigure}[b]{0.4\textwidth}
\centering
\includegraphics[width=\textwidth]{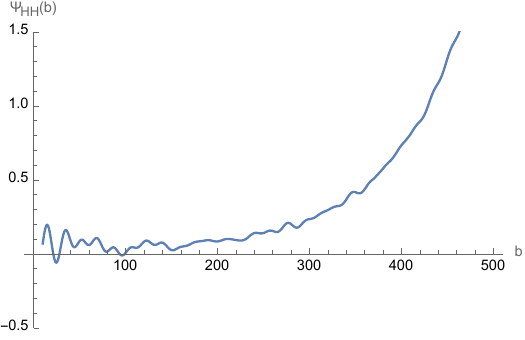}
\caption{Atypical draw of $H_0$}
    \label{b}
    \end{subfigure}
    \caption{We plot the Hartle-Hawking wave function $\hh(b)$ for a typical and an atypical draw of the ensemble of Hamiltonians. At large $b$ the atypical draw better approximates the bulk gravity answer. See footnote \ref{footnote:resummation}.}
    \label{fig:Plot_2}
\end{figure}
Figure \ref{fig:Plot_2} contains plots of the wave function \eqref{eq:HH_dictionary}. In \ref{fig:Plot_2}(\subref{a}) we plot the typical draw of the ensemble, while in \ref{fig:Plot_2}(\subref{b}) we plot an atypical draw.\footnote{Plot \ref{fig:enter-label} and plot \ref{fig:Plot_2}(\subref{a}) are from the same draw of $H_0$. We zoomed into figure \ref{fig:enter-label} to illustrate the erratic nature of the wave function. Here, by ``atypical draw", we mean it has at least one negative eigenvalue, which causes growth in the expectation value of $\mathcal{O}(b)$ \eqref{eqn:matrixoperator_O}. The probability to get a Hamiltonian with a negative eigenvalue  depends on the details of the theory.} It is atypical in the sense that its spectrum contains an eigenvalue that is negative after shifting the edge of the spectrum to take the double scaling limit. This gives rise to growth as a function of $b$ which is the expected bulk gravity answer.\footnote{\label{footnote:resummation}From the boundary perspective, the exact wavefunction \eqref{eqn:matrixoperator_O} will grow with $b$ if we have a negative eigenvalue. A typical eigenvalue of size $\sim-e^{-S_0}$ would give growth $\sim \frac{1}{b} \cosh \lr{\# b e^{-S_0/2}}$. From the bulk perspective, the Weil-Petersson volumes in \eqref{eqn:AveragedHH} also grow as functions of $b$. After doing a full resummation of the genus expansion along with the inclusion of other doubly non-perturbative effects, the theory should see that there are negative eigenvalues in the spectrum. We thus see a qualitative match between the numerical and bulk computations, but these answers do not quantitatively match. This is because the exact answer will be sensitive to the entire probability distribution of negative eigenvalues.
} One surprising lesson is that certain features of the averaged wave functions actually come from behaviors of atypical members of the ensemble.

We would like to emphasize that even though the wave function in a single theory is unique, and as we have argued trivial, the coarse grained features of the averaged wave function are captured by the wave function in a single theory. In this sense even though there is a unique state it seems to capture features of a sum over geometries in the averaged theory.

\section{Discussion}
\label{sec:discussion}

In this paper we have studied closed universes in holographic theories of quantum gravity. We have shown that in a single theory, factorization of boundary partition functions implies that there is unique closed universe state given by the Hartle-Hawking wave function. We have argued that for holographic theories that are chaotic this wave function will be highly erratic. 

We have explicitly shown these properties in a simple two dimensional model of quantum gravity given by JT gravity. One surprising conclusion is that the wavefunction of closed universes is encoded in a highly non-trivial way in the boundary data of the dual theory.  We end with some open questions and comments.

\paragraph{Ensemble averaging.}
We see that ensemble averaging plays a crucial role in obtaining smooth and rich semi-classical physics. Is it necessary? If it is, it will be quite different from the standard gauge-gravity duality for spacetimes with asymptotic boundaries. While some low dimensional bulk theories are known to be dual to an ensemble average of boundary theories, how should one do ensemble averaging in higher dimensional quantum mechanical theories like super Yang-Mills? Another way to ask this is, how do smooth and semi-classical wave functions emerge for theories for which there is no obvious ensemble averaging procedure?

\paragraph{Observables in closed universes. }

It is an interesting open problem to find clear observables for closed universes. Since the wave function in a single theory is unique, and all input states are secretly equivalent, it is unclear in what sense $\hh(b)$ contains information. Nevertheless, we showed that if you naively examine $\hh(b)$ for a single theory, it seems to qualitatively reproduce on a coarse grained level some features of the semi-classical wave function in the averaged theory. It thus seems tempting to speculate that one may be able to extract non-trivial physics from $\hh(b)$ even within a single theory.

Ensemble averaging naively appears to significantly improve the situation. We obtain a large number of orthogonal states and we can define some set of observable with these states. However, naively defined observables seem to have large fluctuations due to effects of spacetime wormholes \cite{Usatyuk:2024mzs} when the variance of the observable is calculated, and so averaging ultimately doesn't give a more sensible quantum mechanical description of closed universes. The situation would be improved if the rules were somehow modified to remove wormhole contributions, but this seems unnatural from recent developments in black hole physics. It would be interesting if there was a natural mechanisms to suppress large wormhole fluctuations for closed universes.

\paragraph{Expanding closed universes.} In this work we have taken the approach of starting from a well defined holographic theory, and studied closed universes within the theory. The simplest examples of closed universes we find with this approach are AdS big bang/big crunch cosmologies \cite{Maldacena:2004rf,VanRaamsdonk:2020tlr}. It would be interesting if there were examples of holographic theories where the closed universe had other behavior, such as expanding eternally like de Sitter space \cite{Kachru:2003aw}. 

\paragraph{Factorization for finite boundary conditions.} In this work it was important that the bulk path integral factorizes when evaluated with multiple boundary conditions. This includes both asymptotic and finite boundaries. In the case of asymptotic boundaries, this must be true since we are evaluating boundary CFT partition functions which are well defined quantities that factorize. However, for boundary conditions that are not asymptotic, such as finite boundaries used for the evaluation of the Hartle-Hawking wavefunction, the argument for factorization is less clear. It would be desirable if the path integral for finite slices was also non-perturbatively defined in terms of some independent quantum mechanical theory. In the case of JT gravity we found this is precisely the case; the geodesic boundary is independently defined by the data of the boundary quantum mechanics \eqref{eqn:matrixoperator_O}, and so must factorize.

Does this property continue to hold in higher dimensional theories such as the bulk dual to $\mathcal{N}=4$ super Yang-Mills? One appealing approach is that of Cauchy Slice Holography \cite{Araujo-Regado:2022gvw,Araujo-Regado:2022jpj}, where the quantum gravity wave function on a non-asymptotic slice can be defined in terms of a $T\overline{T}$ flowed partition function of the boundary theory evaluated on the slice.\footnote{This is currently being investigated in the case of JT gravity, where the quantum theory at finite cutoff is a deformation of the Schwarzian theory\cite{JTCSH}.} This would provide an independent quantum mechanical definition of the wave function on non-asymptotic slices, and would explain the uniqueness and factorization properties of such boundary conditions.

\section*{Acknowledgment}
We are grateful to Douglas Stanford for explaining the bulk origin of negative eigenvalues, and Don Marolf for helpful comments on the draft. M.U. was supported in part by grant NSF PHY-2309135 to the Kavli Institute for Theoretical Physics (KITP), and by a grant from the Simons Foundation (Grant Number 994312, DG). Y.Z. was supported in part by the National Science Foundation under Grant No. NSF PHY-1748958 and by a grant from the Simons Foundation (815727, LB).

\appendix
\section{Review of factorization in deformed JT gravity dual to a fixed Hamiltonian}
\label{app:factorization}

 Let us first briefly explain how the factorization interactions in \eqref{eq:action_deformed} leads to a cancellation of all wormholes in the gravity path integral, and results in factorization between disconnected boundaries. Let's first compute the leading contribution to the connected component of the product of two thermal partition functions
\be
Z_{\t{conn}}(\beta_1, \beta_2) = \raisebox{-.75cm}{\includegraphics[width=8cm]{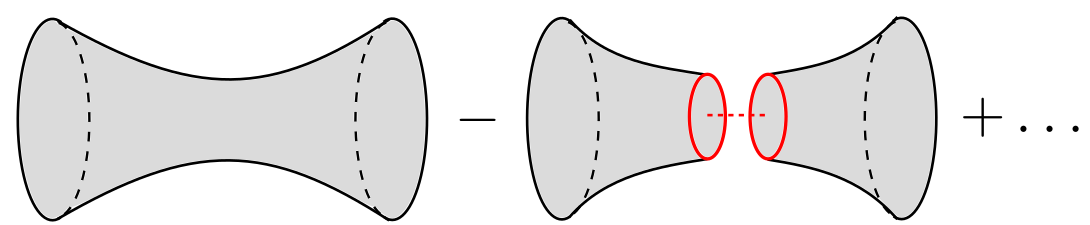}}
\ee
The double trumpet contribution is cancelled out by a single factorizing interaction. We see that 
Order by order the factorizing interaction will give contributions that subtract off the wormhole contributions.\footnote{The basic idea behind the interaction is that all hyperbolic surfaces can be decomposed into geometries where the asymptotic boundary is separated from the rest of the geometry by a geodesic $b$. The introduction of $-\frac{1}{2}\int_0^\infty d b b \mathcal{O}(b) \mathcal{O}(b)$ introduces additional linked geodesic boundaries that cancel out any corresponding wormhole.} In \cite{Blommaert:2021fob} it was shown that these cancellations work to all orders in the genus expansion. The only geometries that are not cancelled are those which do not have a geodesic circle somewhere on the geometry, with the only such geometry being the disk.

We now discuss the effect of the interaction vertex that gives a discrete spectrum. Let us calculate the disk amplitude
\be \label{eqn:Zdisk_factorizing}
Z(\beta) = \raisebox{-.75cm}{\includegraphics[width=5cm]{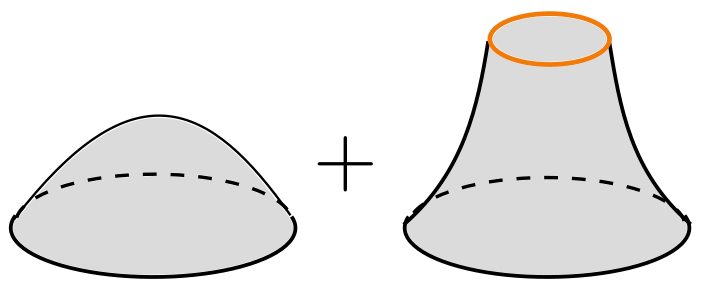}}
\ee
The only surviving geometries to all orders in the genus expansion are the disk and the trumpet weighed by $Z_{H_0}(b)$.\footnote{As a reminder, self-interactions for the vertices have already been re-summed so we do not have have a factorizing vertex connect to the $Z_{H_0}(b)$ vertex to cancel out this trumpet.} We can obtain a discrete spectrum corresponding to a boundary Hamiltonian $H_0 = \t{diag} \lr{E_1, \ldots, E_N}$ by choosing
\be
Z_{H_0}(b)=\sum_{i=1}^N \frac{2}{b} \cos \left(b \sqrt{E_i}\right)-\int_0^{\infty} d E \rho_0(E) \frac{2}{b} \cos \left(b \sqrt{E} \right)\,,
\ee
where $\rho_0(E)=\sinh \lr{2\pi \sqrt{E}}$ is the disk density of states of JT gravity, and the subtracted term has the effect of getting rid of the disk contribution in \eqref{eqn:Zdisk_factorizing}. With this choice, we find
\be
Z(\beta) = Z_{\t{disk}}(\beta) + \int_0^\infty d b b Z_{\t{trumpet}} (\beta,b) Z_{H_0}(b) = \Tr e^{-\beta H_0}\,,
\ee
where we have used that $ Z_{\t{trumpet}}(\beta, b) = \frac{1}{\sqrt{\beta}}e^{-\frac{b^2}{4\beta}}$ and $Z_{\t{disk}}(\beta)=\int d E \rho_0(E) e^{-\beta E}$. We thus see that the effect of the $Z_{H_0}(b)$ interaction is to get the bulk path integral to reproduce a theory with a discrete spectrum given by a Hamiltonian $H_0$. Formally, we should send the number of eigenvalues $N \to \infty$.

\bibliographystyle{utphys}
\bibliography{references}
\end{document}